\documentclass[aps,prl,twocolumn,superscriptaddress,showpacs]{revtex4}

\usepackage{graphicx}
\usepackage{dcolumn}
\usepackage{bm}

\begin{document}

\title{Directed transport with real-time steering and drifts along
  pre-designed paths using a Brownian motor}

\author{H. Hagman}
\email{henning.hagman@physics.umu.se}
\author{M. Zelan}
\author{C. M. Dion}
\affiliation{Department of Physics, Ume{\aa} University, SE-901 87 Ume{\aa}, Sweden.}
\author{A. Kastberg}
\affiliation{Laboratoire de Physique de la Mati\`{e}re Condens\'{e}e, CNRS UMR
  6622,Universit\'{e} de Nice-Sophia Antipolis, Parc Valrose, F-06108 Nice
  Cedex 2, France}

\date{\today}

\begin{abstract}
  We have realized real-time steering of the directed transport in a
  Brownian motor based on cold atoms in optical lattices, and
  demonstrate drifts along pre-designed paths.  The transport is
  induced by spatiotemporal asymmetries in the system, where we can
  control the spatial part, and we show that the response to changes
  in asymmetry is very fast.  In addition to the directional steering,
  a real-time control of the magnitude of the average drift velocity
  and an on/off switching of the motor are also demonstrated.  We use
  a non-invasive real-time detection of the transport, enabling a
  feedback control of the system.
\end{abstract}


\maketitle

\section{Introduction}

Many microscopic systems have a noisy dynamic governed by thermal
fluctuations, making their control and theoretical treatment
complicated. However, Brownian motors and Brownian ratchets take
advantage of this noise as they convert random fluctuations into
directed motion in the absence of bias
forces~\cite{R.DeanAstumian05091997,BM2,BM1,RevModPhys.81.387}. This
makes them an interesting subject of statistical physics, and they are
the driving mechanisms of a variety of biological
motors~\cite{R.DeanAstumian05091997,RevModPhys.69.1269}, ranging from
inter-cell transport and virus translocation to muscle
contraction~\cite{Alberts1998291,RonaldDVale04072000,Vale2003467,Oster2003114}. Inspired
by these biological machines, several proposals exist to utilize the
principles of Brownian motors to power up future
nano-technology~\cite{MartinG.L.vandenHeuvel07202007,Man-MadeNanomachines}.
Many actual Brownian motors are relatively large and complex systems,
therefore models with more comprehensive and controllable designs are
needed~\cite{RevModPhys.69.1269,R.DeanAstumian05091997,RevModPhys.81.387}. Beside
the naturally occurring biological motors a number of artificial
Brownian motors and ratchets have been realized, such as cold atom
Brownian motors and
ratchets~\cite{PhysRevLett.82.851,OurBM1,PhysRevLett.98.073002,PhysRevA.81.013416}
and quantum
ratchets~\cite{H.Linke12171999,PhysRevLett.90.056802,TobiasSalger11272009}.

For a directed motion to be induced in any of these systems two
requirements have to be fulfilled: the spatial and/or temporal
symmetries have to be broken, in accordance with the Curie
principle~\cite{curie}, and the system has to be brought out of
thermal equilibrium, in agreement with the second law of
thermodynamics~\cite{rat_praw}.  Fulfilling these requirements is
sufficient to induce drifts, which is well established, and a number
of different types of systems have been
demonstrated~\cite{RevModPhys.81.387,Spec_appl_phys_A}.  Reversals of
the induced drift have also been demonstrated,
\emph{e.g.},~\cite{cont_multi_rev}, but the response to an asymmetry
changing in real time, and in three dimensions, hasn't been fully
investigated. All these aspects can be addressed by Brownian motor
systems realized with laser cooled atoms interacting with
three-dimensional (3D) dissipative optical
lattices~\cite{Grynberg2001335}. These make a promising testbed for
fundamental studies of the underlying mechanisms in
play~\cite{RevModPhys.81.387}, and they can be utilized as proof of
principle experiments, or as experimental demonstrations of the
feasibility of Brownian motors.

We here demonstrate a Brownian motor, realized with cold atoms
interacting with two optical lattices, where the asymmetry, and hence
the average drift of the atoms, is controllable in real time in
3D. The optical lattices are individually symmetric, and the required
asymmetry originates instead from a combination of a relative
translation of, and unequal transfer rates between, the optical
lattices. This gives a flexible setup where the rectification can be
controlled via the relative translation of the lattice potentials
\cite{OurBM1}.  We here introduce an external real-time control of the
translation of the lattices, by use of electro-optical modulators,
along with real-time, non-destructive measurement by fluorescence
imaging.  This allows us to investigate the response of the system to
a changing asymmetry, as well as the possibility to utilize these
asymmetry changes for external real-time steering of the induced
drift, and for inducing of drifts along pre-designed paths.

\section{Working principle}

\begin{figure}[htbp]
\centering
\includegraphics[width=\columnwidth]{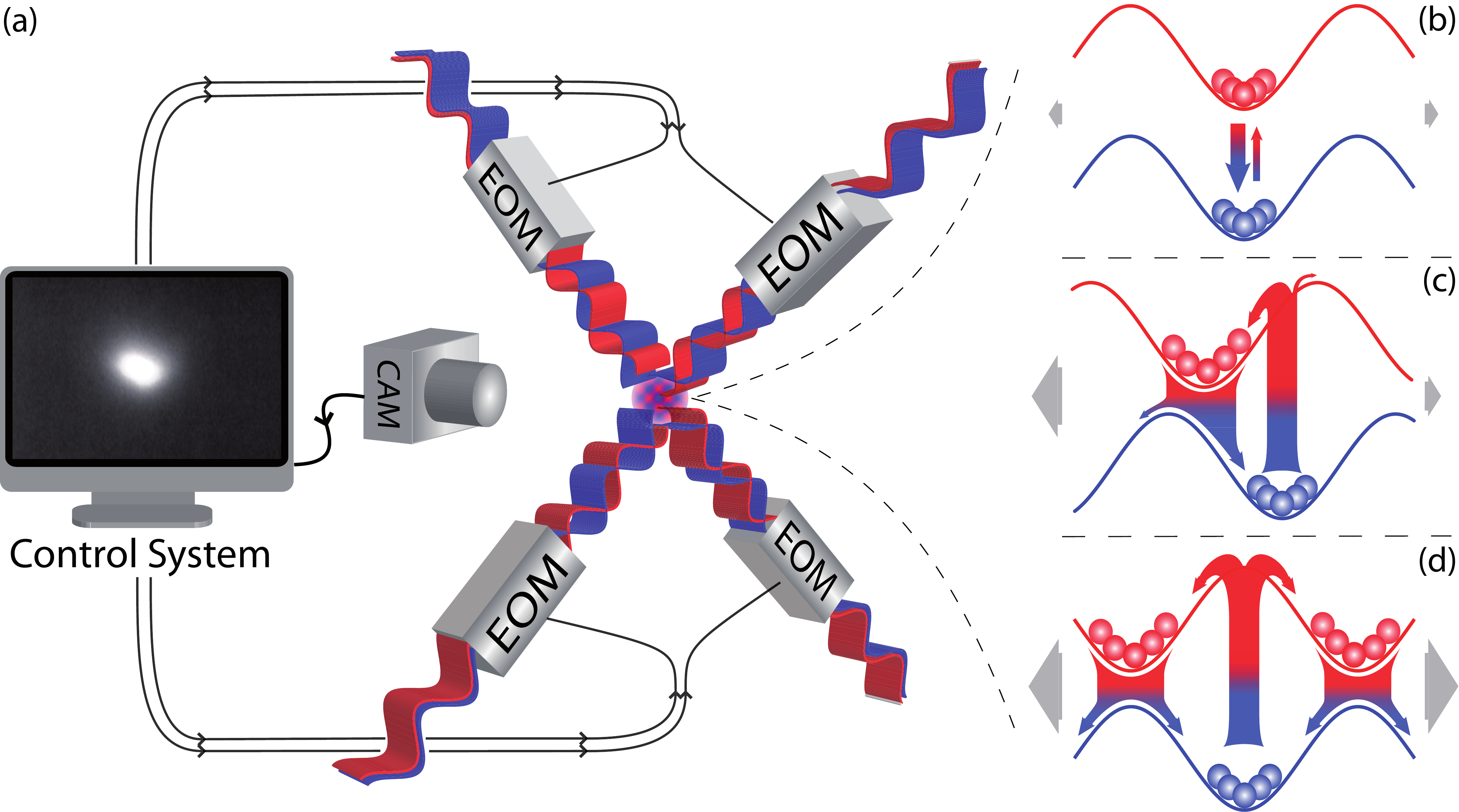}
\caption{(Color online) (a) Schematic representation of the key
  elements of the experimental set-up. The two optical lattices are
  constructed from the interference patterns from two superposed,
  four-laser-beam configurations. In each arm a computer controlled
  EOM is located, changing the relative phase $\varphi$. The
  fluorescence of the atoms is monitored by the control system through
  a CCD camera. (b,c,d): 1D representation of the atoms in the
  long-lived (lower, blue) and the short-lived (upper, red) optical
  lattices.  In each potential an inherent friction and diffusion is
  present. The vertical arrows indicate the transfer between the
  potentials, and the horizontal the total diffusion. (b) $\varphi =
  0$, the system is symmetric and the particles are mostly located
  near the bottom of the wells, undisturbed by the inter-potential
  transfer. No drift is induced. (c) $\varphi = 2/3\pi$, the transfer
  adds energy to the system, the symmetry is broken, and a drift to
  the left is induced. (d) $\varphi = \pi$, the overall symmetry is
  restored, and the added energy only gives an increased, unbiased
  diffusion.}
\label{idea}
\end{figure}

To qualitatively understand the induced drift dependence on the
translation between the potentials, a simple 1D model is used
\cite{bm:sanchez-palencia04}, see Fig.~\ref{idea}.  Consider a classical
Brownian particle situated in either of two symmetric and periodic
potentials (S and L). Both potentials have an inherent friction and
Brownian diffusion, and are coupled with unequal transfer rates.  The
particle randomly switches potential, spending a longer time in
potential L.  When the potentials are in phase, the particle will
mostly be located near the bottom of a well, undisturbed by the
inter-potential transfer, see Fig.~\ref{idea}b.  The system is symmetric and
no drift is induced. However, if one potential is translated with
respect to the other (a relative phase shift $\varphi$), the
spatio-temporal symmetry is broken, and the passage from one potential
to the other, on average, adds energy to the
particle~\cite{PhysRevE.68.021906}. This gain in energy, together with
the relative translation, leads to an increased diffusion, which due
to the total asymmetry is biased (Fig.~\ref{idea}c), with maximum drift
occurring around $\varphi = \pm 2\pi/3$. A translation of half a
period only increases the unbiased diffusion since the symmetry is
restored, see Fig.~\ref{idea}d.  Although it contains all the essential
ingredients of the Brownian motor, this model is an simplification of
the experimental system~\cite{OurBM1,0295-5075-81-3-33001}.  For
cesium atoms, each of the two potentials is a manifold of potentials
of different amplitudes~\cite{Grynberg2001335}, of which one dominates
the dynamics. Moreover, the damping force applied to an atom, and the
transfer rates between the potentials, are position and velocity
dependent~\cite{DOL,Grynberg2001335}.

\section{Experimental setup}

Detailed descriptions of the experimental setup are given in
\cite{DOL,OurBM1}.  In short, about $10^8$ cesium atoms are
accumulated and cooled to around $5\ \mu$K in a magneto-optical trap
\cite{Metcalf:03}. The latter is then switched off, and two optical
lattices~\cite{Grynberg2001335} are superposed on the atoms. These
optical lattices are 3D light-shift potentials, created in the
interference pattern of laser beams, and are dissipative
\cite{Grynberg2001335}, that is, they are constructed from light
fields which are tuned sufficiently close to an atomic resonance for
light scattering to be important, providing a source of noise as well
as an inherent friction due to laser cooling
\cite{Grynberg2001335,DOL}.  Moreover, the two optical lattices are
spatially overlapped, have the same periodicity, and an atom interacts
with either depending on its hyperfine state.  The incoherent light
scattering provides a route between the two optical lattices, via a
manifold of short-lived excited states. The transfer rates are highly
unequal, which results in one short-lived and one long-lived optical
lattice (denoted S and L, respectively).  The potential depths
correspond to around $100\ \mu$K in lattice S and $200\ \mu$K in
lattice L, while the kinetic temperature of the atoms is around $10\
\mu$K. The kinetic temperature has a dependence on the relative
translation of the potentials~\cite{DOL}, and can therefore be used as
a measurement of this translation, or for monitoring the relaxation to
steady state after a change in the system.

In contrast to previous work, the relative translation of the optical
lattices is now controlled with electro-optical phase modulators
(EOMs).  That is, the optical lattice configuration consists of four
arms, each with two laser beams with orthogonal polarizations. An EOM
is placed in each arm to control the relative phase of the two beams.
The crystals inside the EOMs have one electro-optical active axis and
one passive axis.  Along the active axis, the index of refraction is
dependent on an externally applied voltage allowing the external
control of the optical path length, while along the passive axis it
remains unchanged.  By controlling the relative optical phase of each
arm, the relative spatial phase of the resulting potentials can be
controlled as well. After the EOMs, but before the arms start to
interfere, the polarization of the two beams in each arm are turned to
the same direction.  The setup is also modified to image the atoms
through the inherent fluorescence in the optical lattices (due to
light scattering), such that the detection doesn't interfere with the
system.  The current experimental setup is illustrated schematically
in Fig.~\ref{idea}a.

\section{Results}
\begin{figure}[htbp]
\centering
\includegraphics[width=\columnwidth]{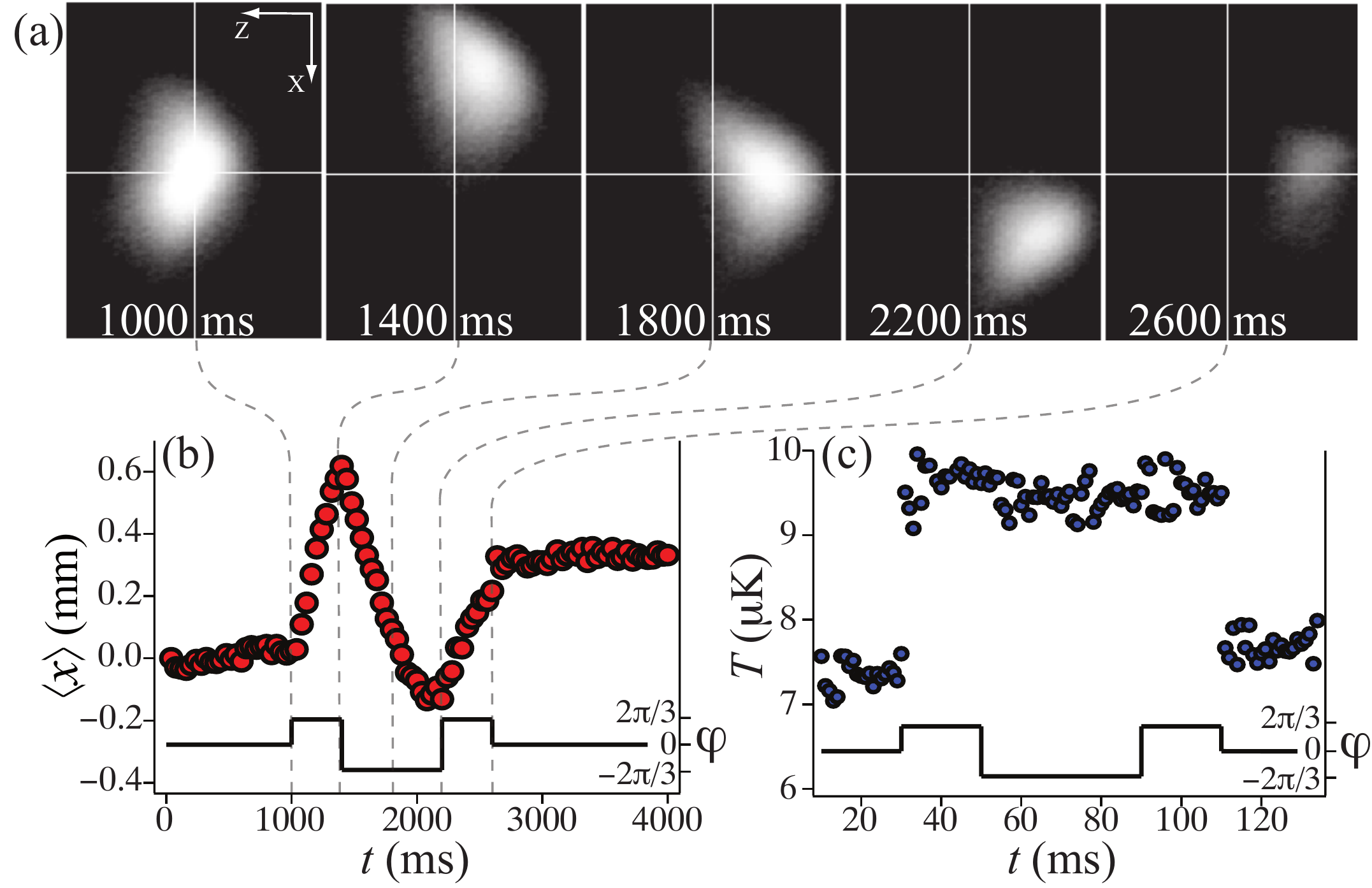}
\caption{(Color online) External steering of the induced drift in 1D. The translation
  between the potentials, ($\varphi_x,\varphi_y,\varphi_z$), are
  changed according to the sequence $(0,0,0) \rightarrow (2\pi/3,0,0 )
  \rightarrow (-2\pi/3,0,0 ) \rightarrow (2\pi/3,0,0 ) \rightarrow
  (0,0,0 )$. (a) Selected images of the atomic cloud in the
  xz-plane, with $z$ directed upwards in the experiment. (b) Time
  evolution of the center-of-mass position along $x$. (c) Time
  evolution of the temperature for the same sequence of translational
  shifts, but on a shorter time scale. In the lower part of (b) and
  (c) the phase sequences are plotted.}
\label{im}
\end{figure}

%
%
To experimentally investigate the real-time response of the Brownian
motor to changes in the relative translation of the optical
potentials, ($\varphi_x,\varphi_y,\varphi_z$), we start with
translations in 1D. This is done in five steps: $(0, 0, 0)$
$\rightarrow$ $(2\pi/3, 0, 0)$ $\rightarrow$ $(-2\pi/3, 0, 0)$
$\rightarrow$ $(2\pi/3, 0, 0)$ $\rightarrow$ $(0, 0, 0)$, see
Fig.~\ref{im}a where a selection of images, representing the extreme
points of the trajectory, are presented. The whole trajectory is
available online as movie S1~\cite{ms1} The cloud is imaged in the
xz-plane, with $z$ being vertical in the experiment, but horizontal
in the images. In the figure, the expected back and forth trajectory
in $x$ is evident. Beside the back and forth motion, a small downwards
drift is also present, since the optical lattices can't fully support
the atomic cloud against gravity~\cite{grav}. In principle this effect
could be canceled by an imbalanced radiation pressure, or by an
appropriate choice of $\varphi_z$.  In Fig.~\ref{im}b, the time evolution
of the atomic cloud's center of mass along $x$ is presented. The drift
velocities are constant for a fixed translation, and the change in
direction when the potentials are translated appears to be very
fast. To further investigate this, we have repeated the same set of
relative potential translations, but on a shorter time scale, and
measured the temperature using a (destructive) time-of-flight
technique~\cite{hagman:083109}. Figure~\ref{im}c shows that the reaction time
for the atoms to reach the new steady state is less than 1~ms,
\emph{i.e.}, is not resolvable, given the time resolution and
uncertainty of our control and detection systems.

\begin{figure}[htbp]
\centering
\includegraphics[width=\columnwidth]{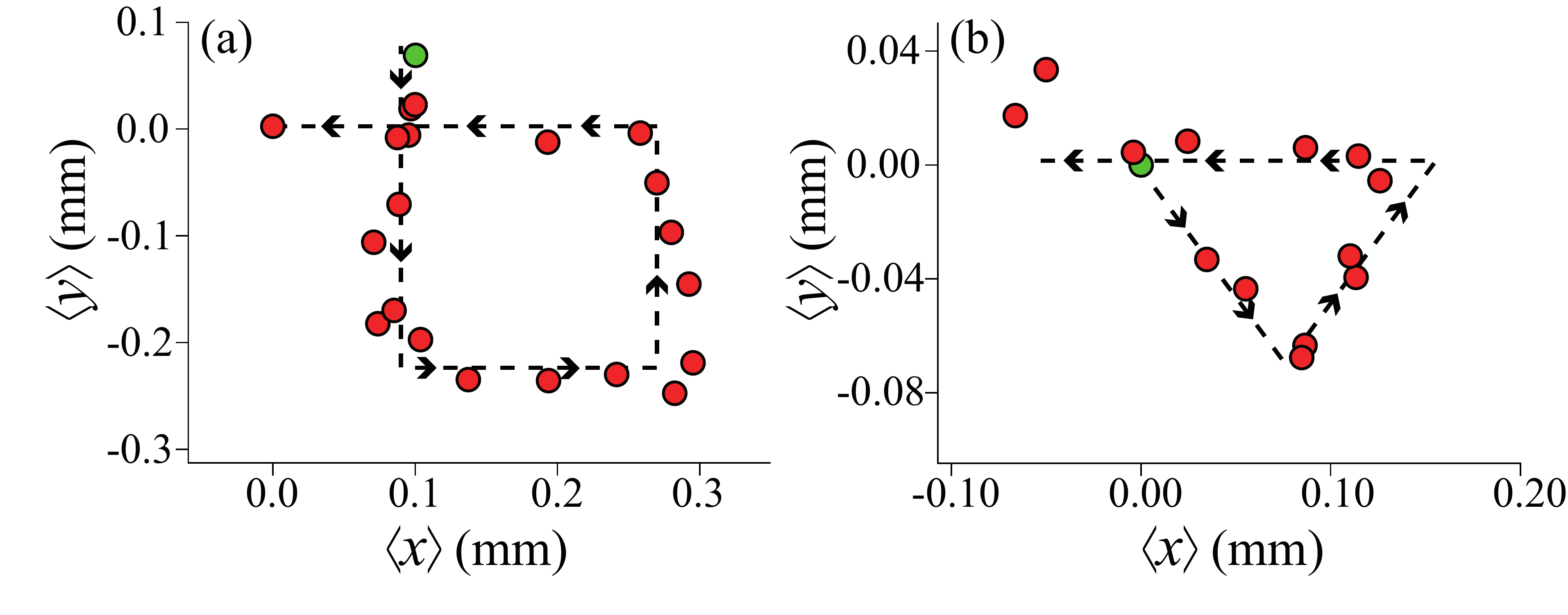}
\caption{(Color online) Drifts along pre-designed closed paths: (a)
  square; (b) triangle. The trajectory of the center of mass is marked
  with filled circles, starting from the green (light gray) one, with
  the anticipated path indicated by a dashed line. The time interval
  between each marker is roughly 75 ms.}
\label{square}
\end{figure}

The relative translation of the potentials can also be altered in
other directions, making it possible to move the atoms along arbitrary
pre-designed paths, including closed figures. We first make the atoms
move along paths with straight angles, as the relative phases of the
potentials are changed in four steps: $(2\pi/3, 0, 0)$ $\rightarrow$
$(0, -2\pi/3, 0)$ $\rightarrow$ $(-2\pi/3, 0, 0)$ $\rightarrow$ $(0,
2\pi/3, 0)$.  The interval between each shift is around
300~ms. Figure~\ref{square}a (and movie S2~\cite{ms1}) show the
expected square path.

Since drifts in arbitrary directions are achievable, any geometrical
figure can in principle be realized. Figure~\ref{square}b (and movie S3
\cite{ms1}) demonstrates the guiding of the atoms along a triangular
path, achieved by relative phase settings of $(2\pi/3, 0, 0)$
$\rightarrow$ $(-\pi/3, -\pi/3, 0)$ $\rightarrow$ $(-\pi/3, \pi/3,
0)$.  Note that the phase settings for off-axis drifts become
non-trivial due a coupling between the dimensions of the lattice
topography~\cite{0295-5075-81-3-33001}.

%
All data where taken for relative translations of the potentials that optimize the magnitude of the drift.  It can however be made arbitrary small~\cite{OurBM1,0295-5075-81-3-33001}, giving us not only real-time control of the direction, but also of the speed of the atoms. In addition, the directional shifts do not need to be discrete, as the electro-optical modulators can be controlled continuously.  Smooth velocity changes are thus achievable, allowing for curved paths. 

With the exception of Fig.~\ref{im}c, all the results presented are
recorded by imaging the inherent fluorescence of the atomic cloud in
the optical lattices. Therefore the detection does not interfere with
the system, enabling a real-time analysis of the position of the
atomic cloud.  This opens up the possibility of implementing a
feedback loop, that is, an autonomous system can be created, where the
atomic cloud's current position and velocity determines the system's
coming actions.

\section{Conclusion}

In summary, we have realized a Brownian motor with a real-time
external steering in 3D, and demonstrated drifts along pre-designed
paths. The directional shifts of the average velocity was shown to be
fast ($<1$ ms), with induced average velocities are up to a few mm/s,
while the lifetime of the atoms in the optical lattice is of the order
of seconds. This gives ample time to induce drifts with several
resolvable directional shifts, see movie S4~\cite{ms1}. On a more
general scale, a typical velocity of 1 mm/s means that each atom, on
average, moves over 1600 potential wells per second.  We showed that
the induced drifts can be detected in real time and non-destructively,
opening possibilities for feedback control of the induced drift. The
system could hence be controlled externally in real time by,
\emph{e.g.}, a joystick, by pre-determine protocols inducing drifts
along designed paths, or via a position feedback that enables the
system to autonomously steer itself. The feedback could be used,
\emph{e.g.}, to stabilize drifts or for the system to find its own way
through a maze. Real-time steering, drifts along pre-designed paths,
and feedback controlled drifts are all important for the creation of
useful future applications of Brownian
motors~\cite{MartinG.L.vandenHeuvel07202007,Man-MadeNanomachines,Hess:chem}. Our
system may also function as a model system that can be used for
fundamental studies of the properties and feasibility of Brownian
motors.

\begin{acknowledgments}
  We thank S. Jonsell and G. Labaigt for helpful discussions.  This
  project was supported by the Swedish Research Council, Knut \& Alice
  Wallenbergs Stiftelse, Carl Trygger Stiftelse, Kempe Stiftelserna,
  and Ume{\aa} University.
\end{acknowledgments}


\end{document}